\documentclass[preprint]{aastex}
\usepackage{natbib}
\usepackage{ulem}
\shorttitle{Cosmic rays and subshock instabilities}
\shortauthors{Stroman et al.}

\begin{document}

\title{COULD COSMIC RAYS AFFECT INSTABILITIES IN THE TRANSITION LAYER OF
NONRELATIVISTIC COLLISIONLESS SHOCKS?}

\author{Thomas Stroman\altaffilmark{1}}
\affil{Department of Physics and Astronomy, Iowa State University,
    Ames, IA 50011, USA}

\author{Martin Pohl}
\affil{Institut f\"ur Physik und Astronomie, Universit\"at Potsdam, D-14476 Potsdam-Golm, Germany \\
and DESY, D-15738 Zeuthen, Germany}

\author{Jacek Niemiec}
\affil{Institute of Nuclear Physics PAN, ul. Radzikowskiego 152, PL-31-342 Krak\'{o}w, Poland}

\author{Antoine Bret}
\affil{ETSI Industriales, Universidad de Castilla-La Mancha, E-13071 Ciudad Real, Spain \\
  and Instituto de Investigaciones Energ\'eticas y Aplicaciones Industriales,
  Campus Universitario de Ciudad Real, E-13071 Ciudad Real, Spain
}
\altaffiltext{1}{Now at Department of Physics and Astronomy, University of Utah, 
Salt Lake City, UT 84112, USA}
\begin{abstract}
There is an observational correlation between astrophysical shocks and
non-thermal particle distributions extending to high energies. 
As a first step toward investigating
the possible feedback of these particles on the shock at the microscopic level,
we perform particle-in-cell (PIC) simulations of a simplified environment consisting
of uniform, interpenetrating plasmas, both with and without an additional population of
cosmic rays. We vary the relative density of the counterstreaming plasmas, the
strength of a homogeneous parallel magnetic field, and the energy density in cosmic
rays. We compare the early development of the unstable spectrum for selected configurations
without cosmic rays to the growth rates predicted from linear theory, for assurance that the
system is well represented by the PIC technique. Within the parameter space explored, we
do not detect an unambiguous signature of any cosmic-ray-induced effects on the microscopic
instabilities that govern the formation of a shock. We demonstrate that an overly coarse
distribution of energetic particles can artificially alter the statistical 
noise that produces the perturbative seeds of instabilities, and that such effects
can be mitigated by increasing the density of computational particles.
\end{abstract}
\keywords{cosmic rays, instabilities, plasmas, shock waves}

\section{Introduction}\label{sec:p3int}
Shocks form in a wide variety of astrophysical environments, from planetary bow
shocks in the heliosphere to colliding clusters of galaxies. The presence of
nonthermal particle populations in some of these environments 
is inferred from radio, X-ray, and $\gamma$-ray observations 
\citep{2008ARA&A..46...89R},
and a number of theoretical mechanisms link strong shocks (such as those found
in young supernova remnants) to
particle acceleration processes \citep[e.g.,][]{1983RPPh...46..973D,1983ApJ...270..537W}. 
An understanding of the
nonlinear coupling between the shock, the energetic particles, and
the spectrum of any excited waves is essential to the proper interpretation
of the observational signatures of shocked systems as well as to a
correct understanding of the origin of cosmic rays.

In most cases the density of the shocked medium is sufficiently low for collisions
between particles to be infrequent; such collisionless shocks are mediated
through collective electromagnetic effects. The thickness of the shock
transition region and the nature of the microphysics occurring therein may
play a key role in the injection efficiency of thermal particles into the acceleration 
processes thought to accelerate cosmic rays to high energy 
\citep{1978MNRAS.182..147B,1978MNRAS.182..443B,2002PhPl....9.4293S}. 
In addition,
efficient particle acceleration associated with some shock environments
raises the question of the extent to which a significant cosmic-ray
contribution to the local energy density can influence the structure
of the shock front itself. This may be of particular importance in
starburst galaxies, where massive stars and frequent supernovae
increase the availability of sites for particle acceleration. The emission of
TeV $\gamma$-rays from some nearby starburst galaxies
suggests an abundance of cosmic rays that significantly exceeds the
values observed locally \citep{2009Natur.462..770V}.

A number of mechanisms exist whereby cosmic rays may play a role in
shaping the shock and its environment.
A significant cosmic-ray contribution to the pressure upstream of a shock is 
understood to result in substantial modification to the shock environment \citep{1984ApJ...277..429E}; 
this effect is expected to produce an effective shock compression ratio that is perceived
as being larger by cosmic rays of higher energy, possibly hardening the local
cosmic-ray source spectrum \citep{2001RPPh...64..429M}. Under some conditions,
the current carried by
cosmic-ray ions diffusing ahead of the shock may excite instabilities in the
upstream interstellar medium that lead to large-amplitude magnetic turbulence
\citep{2004MNRAS.353..550B,2011MNRAS.417.1148L,2011ApJ...736..157R,2011ApJ...738...93N} or 
other substantial alterations to the upstream environment, 
such as the heating of thermal electrons \citep[e.g.,][]{2008ApJ...684..348R}. However,
these effects occur on spatial scales much larger than the thickness of the subshock and
may operate quite independently of the processes occurring therein.

In this paper, we turn our attention to those processes within the subshock.
In order to quantify or constrain the effect of a ``spectator'' population
of cosmic rays (that is, any cosmic rays present 
that have already been accelerated, locally or elsewhere) 
on the formation and evolution of nonrelativistic 
collisionless subshocks, we
design simulations to explore the initial linear and subsequent nonlinear
growth of instabilities that 
contribute to shaping the shock transition region.
We restrict our focus to the interpenetration layer, in which two 
counterstreaming plasma shells
overlap in space. This is fertile ground for the growth of
well-known and well-studied instabilities 
\citep{2003ApJ...596L.121S,2004ApJ...608L..13F,2009ApJ...699..990B},
and the two-plasma system will respond accordingly. We then repeat
our simulations with an added cosmic-ray component, 
a plasma consisting of highly relativistic particles, so we may compare
the systems at both the early and late stages of their evolution and identify
whether the presence of the cosmic rays has any effect.

The scale of the instabilities dictates the computational approach to modeling them.
Hydrodynamical
models of shocks are incapable of accurately resolving length scales 
much smaller than a particle mean free path and must therefore approximate
subshocks as
discontinuities, which is inadequate for our purposes. 
A self-consistent kinetic approach is necessary for a more complete
exploration of the relevant small-scale physical 
processes, but simulating the entire subshock thickness is prohibitively costly. 
We elect a simulation that represents only
a homogeneous portion of the subshock interior, and therefore we
cannot account for a number of instabilities that arise from spatial gradients or
particle reflection. Here we use 
particle-in-cell (PIC) simulations in two spatial dimensions 
to model the interaction of counterstreaming plasma
flows in the presence of a third, hot plasma of energetic particles. The
PIC technique is a (particle-based) kinetic approach to
modeling the self-consistent evolution of an arbitrary distribution of charged
particles and electromagnetic fields and thus is well suited to the nonlinear development
of unstable plasma systems. In
order to isolate the subshock instabilities from larger-scale spatial effects
such as those arising from systematic charge separation, all
distribution functions are initially homogeneous with respect to position.
We consider the simplified case in which two electron--ion plasmas,
not necessarily of equal density, move nonrelativistically through each
other. This movement may also be parallel to a uniform magnetic field. 
We first observe the
evolution of the drift velocities, field amplitudes, particle distributions,
and wave spectra in the case when cosmic rays are negligible.
We then include cosmic rays whose energy density is an order of magnitude below the
kinetic energy in the bulk flow of the plasmas. We also explore the case
of an artificially high energy density in cosmic rays, 
exceeding the value of equipartition with the bulk flows, to gain insight into
effects that may be too small to arise in the case intended to represent
a more realistic environment but that may nevertheless appear in
regions where energetic particles are unusually abundant. However,
our results suggest that even a high abundance of cosmic rays is not
sufficient on its own to produce a significant deviation from the usual
evolution of the instabilities shaping a shock.

\section{Objectives and approach}
To constrain the extent to which cosmic rays might influence the physics of the
shock transition layer, we perform a series of simulations. Using the benchmarks of system behavior outlined
in Section \ref{ssec:p3benchmarks}, we characterize the response of the 
counterstreaming-plasma system to the cascade of instabilities arising from
interactions among the various plasma components in the absence of cosmic rays. 
We then repeat the simulations with cosmic rays present, so as to facilitate
the side-by-side comparison of the various benchmarks. The
parameter choices for the systems that we explore in this manner are
described in detail in Section \ref{ssec:p3config}, and our simulation model
and its implementation are described in Section \ref{ssec:p3setup}. Then,
a test
of selected parameter configurations against the predictions from analytical
studies of related systems is described in Section \ref{ssec:p3comp}. Following
a brief discussion, we conclude in Section \ref{sec:p3conc}.

\subsection{Parameter-space configurations}\label{ssec:p3config}
The physical system under consideration is modeled as one  
homogeneous electron--ion plasma moving relative to another. These plasmas
may in general have different densities.
In addition, a uniform magnetic field may be aligned with the flow direction.
Finally, a population of cosmic rays may be present, at rest in bulk in the
center-of-momentum frame of the two plasmas. This is the reference frame
chosen for the simulation, so the two plasmas flow in opposite directions;
we adopt the nomenclature of ``stream'' and ``counterstream'' to distinguish
between them. By our convention, the stream's velocity is in the $-x$-direction, 
antiparallel to the guiding magnetic field when one is present.

Our primary simulations are sensitive to variations in three parameters:
the density ratio between the stream and the counterstream, the strength of
the guiding magnetic field, and the energy density in the cosmic rays.

The inter-plasma density ratio $w\equiv n_{s}/n_{\rm cs}$ takes three values and their
respective designations: the ``symmetric'' case $w=1$, the ``intermediate'' case $w=0.3$,
and the ``dilute'' case $w=0.1$. The velocity of the stream is fixed at 
${\bf v_s}=-0.2c {\bf \hat{x}}$,
while the velocity of the counterstream obeys the relation 
$n_{\rm cs}{\bf v_{cs}}+n_s{\bf v_s}=0$; in the
dilute-stream case, the relative flow speed is therefore $0.22c$. Likewise,
the density of the counterstream is fixed, so the stream density alone varies; the
total electron density $n_e$ and thus the electron plasma frequency 
$\omega_{\rm pe}=\sqrt{e^2 n_e/\epsilon_0 m_e}$ (where $\epsilon_0$ is the
vacuum permittivity, and $n_e=n_{s}+n_{\rm cs}$ is the cumulative electron density) 
is largest in the symmetric case and reduced
by a factor $\sqrt{1.1/2}$ in the dilute case.

The magnetic field $B_{0,x}$ may be absent, or present at either of two amplitudes given by
the ratio of the electron cyclotron frequency $\Omega_e= e B_{0,x}/m_e$ 
to the electron plasma frequency, 
$b\equiv \Omega_e/\omega_{\rm pe}$. The ``absent'' magnetic
field refers to $b=B_{0,x}=0$. The values designated ``weak'' and ``strong'' correspond to
$b=0.01$ and $b=0.1$, respectively; in a plasma of electron density of, e.g., $n_e \sim 1$~cm$^{-3}$,
a magnetic field of $\sim 300$~$\mu$G is necessary for $b=0.1$. Since the 
speed of Alfv\'en hydromagnetic waves $v_A\equiv b\,c/\sqrt{1+m_i/m_e}$ 
is at most $0.014c$ for our choice of $m_i/m_e=50$, all of the plasma collisions we consider
have Alfv\'enic Mach numbers significantly larger than unity 
(up to $\infty$ in the case when $b=0$). 
Note that because the electron plasma frequency is not 
independent of the density ratio $w$ in our simulations,
neither is the absolute magnetic-field 
amplitude corresponding to a particular value of $b$.

All simulations include cosmic-ray particles consisting of electrons and ions,
initialized according to an isotropic distribution function and a single speed
(Lorentz factor $\gamma_{\rm CR}=50$)
whose statistical weight $w_{\rm CR}\equiv n_{\rm CR}/n_{s}$
is adjusted to three levels: ``negligible'' when $w_{\rm CR}\gamma_{\rm CR}=10^{-8}$,
``present'' when $w_{\rm CR}\gamma_{\rm CR}=10^{-3}$, and ``abundant'' when $w_{\rm CR}\gamma_{\rm CR}=10$.
Since $w_{\rm CR}$ is defined in terms of the stream density $n_s$, the absolute energy
density in cosmic rays also varies with the density ratio $w$, being 10 times larger in
the symmetric case than the dilute case for each value of $w_{\rm CR}$. Neglecting
the contribution of electrons, the bulk
kinetic energy in the stream plus counterstream is of order 
$n_{\rm cs}m_i v_s^2 w(1+w)/2$, while the cosmic-ray energy density is of order 
$n_{\rm CR}\gamma_{\rm CR}m_i c^2=25w_{\rm CR}\gamma_{\rm CR}w n_{\rm cs}m_i v_s^2$, where we have used the
relation $v_s/c=0.2$. Thus, the ratio of cosmic-ray energy 
density to bulk kinetic energy density is $50 w_{\rm CR}\gamma_{\rm CR}/(1+w)$: considerably
larger than unity for the ``abundant'' case and 
a few percent in the ``present'' case.

\subsection{Simulation setup}\label{ssec:p3setup}
Our simulations employ a modified version of the relativistic electromagnetic
PIC code
TRISTAN \citep{1993cspp.book......M}, updated for parallel use with MPI, operating in 
two spatial dimensions, with three-component
velocity and field vectors (2D3V), 
with periodic boundary conditions. 
The charge-conserving current deposition
routine of \citet{2003CoPhC.156...73U} and the field update algorithm with 
fourth-order accuracy
from \citet{2004JCoPh.201..665G} are the most prominent additions, 
as well as digital filtering of
electric currents to suppress small-scale noise via an iterative smoothing algorithm.

The primary set of simulations, 27 in total, was conducted on a spatial grid
in the $x$--$y$ plane of size $280\lambda_{\rm se}\times 180\lambda_{\rm se}$ (periodic boundary 
conditions in $x$ and $y$, with elongation in the
flow direction $x$), where 
$\lambda_{\rm se}\equiv c/\omega_{\rm pe}=10\Delta$ is the electron skin depth, set
to 10 grid cells of length $\Delta$. The electron plasma frequency $\omega_{\rm pe}$
is determined from the sum of the stream and counterstream electron densities only
and thus depends on $w$ but not on $w_{\rm CR}$. Supplementary high-resolution simulations 
in which
$\lambda_{\rm se}=30\Delta$ were performed on a grid of more cells but representing
a smaller physical region, $128\lambda_{\rm se}\times 96\lambda_{\rm se}$. 
The time step $\delta t$ was chosen such that $\omega_{\rm pe}^{-1}\approx 22\delta t$ 
for the $\lambda_{\rm se}=10\Delta$ simulations, or $\omega_{\rm pe}^{-1}\approx 66\delta t$
for the high-resolution $\lambda_{\rm se}=30\Delta$ simulations.

Six separate particle populations from three plasmas are modeled:
the ``stream'' moving in the $-x$-direction,
the ``counterstream'' moving in the $+x$-direction,
and the energetic particles representing cosmic rays. 
Within each plasma, ions and electrons of charge $\pm e$ and mass ratio
$m_i/m_e=50$ have equal charge density and a common drift velocity so that
the entire setup has no net current and no charge imbalance. 
The artificially low mass ratio expedites the simulations, but may change some modes \citep{2007GeoRL..3414109H}.
Quite a few of these possibly modified modes are not included here, partly because we do not consider
the spatial structure of subshocks, partly because they do not fit onto the computational grid, 
and partly because we do not simulate situations involving an oblique or a perpendicular 
large-scale magnetic field. In any case, the analytical treatment of instabilities presented
in Section \ref{ssec:p3comp} accounts for the small mass ratio, and in that sense
our approach is consistent, at least for the linear phase.

Each cell
in a primary (high-resolution supplementary) simulation is
initialized with a total of 90 (120) computational particles: 
20 stream ions, 20 counterstream ions, 
and 5 (20) cosmic-ray ions; and an electron for each ion. The physical density
of each plasma is manipulated through the assignment of the appropriate 
statistical weights, $w$ and $w_{\rm CR}$, to the various particle species.

The stream and counterstream, viewed from their respective rest frames 
at $v_s=-0.2c\hat{x}$ and $v_{\rm cs}=w\times 0.2 c \hat{x}$, are described
by a Maxwell--Boltzmann distribution in which the electrons' most probable speed
is given by $v_{{\rm th},e}=0.01c$ and the ions are in equilibrium with the electrons.
The cosmic rays, whose rest frame is the simulation frame $v_{\rm CR}=0$, are isotropic
and each is initialized with Lorentz factor $\gamma_{\rm CR}=50$, regardless of whether it is an
ion or an electron.
Placing the cosmic-ray population at rest in the center-of-momentum frame minimizes streaming in the collision zone,
and therefore reduces known cosmic-ray-driven instabilities 
\citep{1983A&A...119..274A,2004MNRAS.353..550B,2006PPCF...48.1741R}, which can be 
independently studied \citep[e.g.,][]{2008ApJ...684.1174N,2009MNRAS.397.1402L,2009ApJ...698..445O,2009ApJ...706...38S,2009ApJ...694..626R,2010ApJ...711L.127G}. 
We are interested in whether or not cosmic rays can modify instabilities operating at 
subshocks, and therefore suppress cosmic-ray streaming instabilities .  

\subsection{Behavioral benchmarks}\label{ssec:p3benchmarks}
To provide a basis for comparison among different cosmic-ray densities $w_{\rm CR}$ for
each combination of plasma density ratio $w$ and magnetic-field amplitude $b$, we select
the following attributes of the system for study: the drift velocity
of each particle population, the instantaneous root-mean-square amplitude of 
the parallel and perpendicular components of the electric and magnetic field over the entire
simulation domain, the effective temperature of each stream or counterstream particle species,
and the spectrum of excited wave modes in the magnetic field.

As there is no initial bulk motion in the perpendicular directions, 
only the parallel component $V_x$ of drift velocity is considered. 
The electric field amplitudes are
presented in units of the scaling electric field $E_\omega\equiv\omega_{\rm pe}\,c\,m_e/e$ 
(equivalent to the field at which the electric energy density is half the
electrons' rest-mass energy density, $\epsilon_0 E^2/2=N_e m_e c^2/2$); the magnetic field
multiplied by $c$ is expressible in the same units.

The particle distributions may not remain strictly Maxwellian throughout the
duration of the simulation. As a surrogate for temperature, therefore, 
the mean random kinetic energy of the electrons and ions
of the stream and counterstream is calculated by determining the systematic velocity component
within a $10\Delta\times 10\Delta$ region and eliminating this local bulk motion via an appropriate
Lorentz transformation; the mean post-transformation Lorentz factor $\gamma'$ corresponds only
to the random motion.

Finally, we will explore the effect of cosmic rays on the time evolution of the 
spectral decomposition of the perpendicular (out-of-plane) magnetic field $B_z$
into its spatial Fourier components, both parallel (wave number $k_x=k_\parallel$) 
and perpendicular to the drift (wave number $k_y=k_\perp$). 

\section{Comparison with analytical beam--plasma predictions}\label{ssec:p3comp}
As a test that our simulation results were consistent with theory,
we applied the methods of \citet{2009ApJ...699..990B} to selected stream--counterstream
configurations without cosmic rays. Whether magnetized or not, beam--plasma systems 
(in which a fast, dilute
``beam'' plays a role comparable to that of our stream, with the dense ``plasma''
representing our counterstream) are susceptible to a host of both electrostatic and
electromagnetic instabilities. For flow-aligned wave vectors, electrostatic modes
such as two-stream or Buneman are likely to grow. In the direction normal to the flow,
the filamentation instability (sometimes referred to as ``Weibel'') is usually excited
as well. Finally, modes with wave vectors oriented obliquely are likewise unstable,
so that the unstable spectrum is eventually at least two dimensional.

The full spectrum has been first evaluated solving the exact dispersion equation in
the cold approximation, accounting thus for a guiding magnetic field as well as finite-mass
ions. It turns out that for the present configuration, ions play a very limited role
(in the linear phase) and are not responsible for any unstable modes which would not be
excited if they were infinitely massive. Unlike settings exhibiting nonresonant modes, 
for example
\citep{2004MNRAS.353..550B,2009ApJ...706...38S}, 
where a single proton beam is considered without electrons
moving {\it at the same speed}, we are here dealing with a plasma-shell collision,
where protons and electrons are comoving. As a result, the effect of finite-mass ions is
simply a first-order correction to the electronic spectrum. The dispersion equation
displays the very same branches, and the growth rate is altered by a quantity proportional to
$\mathcal{O}(m_e/m_i)$.

The dispersion equation for arbitrarily oriented modes with the flow along the $x$-axis 
reads \citep{bret:120501}
\begin{equation}
(\omega^2\epsilon_{xx}-k_y^2c^2)(\omega^2\epsilon_{yy}-k_x^2c^2)-
(\omega^2\epsilon_{xy}-k_yk_xc^2)^2 =0,
\end{equation}
where the tensor elements $\epsilon_{\alpha\beta}$ are given by
\begin{equation}\label{eq:epsi_general}
    \epsilon_{\alpha \beta }(\mathbf{k},\omega) = \delta _{\alpha \beta }
    +\sum_j\frac{\omega_{pj}^2}{\omega^2}\int d^3p \, \frac{p_{\alpha }}
{\gamma(\mathbf{p}) }\frac{\partial f_j^0}{\partial p_{\beta }}
+\sum_j\frac{\omega_{pj}^2}{\omega^2}\int d^3p\, 
\frac{p_{\alpha }p_{\beta }}{\gamma(\mathbf{p})^2 }
\frac{\mathbf{k}\cdot \left(\frac{\partial f_j^0}
{\partial \mathbf{p}}\right)}{m_j\omega -\mathbf{k}\cdot \mathbf{p}/\gamma(\mathbf{p}) } ,
\end{equation}
and the sum runs over the species involved in the system. For each species,
the distribution function $f_j^0$ includes both the stream and the counterstream, and
the corresponding plasma frequency $\omega_{pj}$ is calculated using the sum of their
densities. The results in the cold-plasma limit are evaluated 
considering Dirac's delta distribution functions. Though lengthy, calculations are 
straightforward. Setting $k_y=k_\perp=0$ in the equations above allows to derive the 
dispersion equation for flow-aligned modes such as the two-stream ones. Setting 
$k_x=k_\parallel=0$ gives the dispersion equation for the filamentation modes. 
We present analytical results for the symmetric and the diluted beam cases. While such 
expressions can be derived considering either $w=1$ or $w\ll 1$, expressions valid for 
any density ratio $w$ are much more involved (when they exist). This explains why the 
results below for $w=1$ cannot be derived from those with $w\ll 1$.

For flow-aligned wave vectors, the most unstable wave vector $k_{\parallel,m}$ and
maximum growth rate $\gamma_m$ read (with evaluation corresponding to the configuration
displayed in Figure \ref{fig:p3bret})
\begin{eqnarray}
    k_{\parallel,m}\lambda_{\rm se}\frac{\Delta v}{c} & = & \frac{\sqrt{3}\sqrt{1+m_e/m_i}}{\sqrt{2} \Gamma_s^{3/2}}=1.19,\nonumber \\
    \frac{\gamma_m}{\omega_{\rm pe}} &=& \frac{\sqrt{1+m_e/m_i}}{\sqrt{2} \Gamma_s^{3/2}}=0.68,~~\mathrm{symmetric~case},
\end{eqnarray}
and
\begin{eqnarray}
     k_{\parallel,m}\lambda_{\rm se}\frac{\Delta v}{c}&=& \sqrt{1+w}=1.04,~~w\ll 1\nonumber \\
    \frac{\gamma_m}{\omega_{\rm pe}} &=& \frac{\sqrt{3(1+w)}}{2^{4/3}}\frac{w^{1/3}(1+m_e/m_i)^{1/3}}{\Gamma_s}=0.31,~~\mathrm{dilute~case},
\end{eqnarray}
where $\Gamma_s=\left(1-{v_s}^2/c^2\right)^{-1/2}$ is the bulk Lorentz factor of
the stream, which moves in the simulation frame with speed $v_s=0.417c/(1+w)$. 
As seen in Figure \ref{fig:p3bret}, panels (a) and (f), 
oblique modes dominate for the mildly relativistic
conditions of the simulation. The full-spectrum maximum growth rate is thus slightly
larger than the numerical values calculated above for modes propagating along the flow.

For wave vectors normal to the flow, the growth rate reaches its maximum for
$k_\perp=\infty$, with
\begin{equation}
    \frac{\gamma_m}{\omega_{\rm pe}}=2\frac{v_s}{c}\sqrt{\frac{1+m_e/m_i}{\Gamma_s}}=0.41,~~\mathrm{symmetric~case},
\end{equation}
and
\begin{equation}
    \frac{\gamma_m}{\omega_{\rm pe}}= \frac{v_s}{c}\sqrt{\frac{w(1+w)(1+m_e/m_i)}{\Gamma_s}}=0.12,~~\mathrm{dilute~case}.
\end{equation}
These filamentation data have been calculated neglecting the magnetic field,
which is small ($\Omega_e=0.01\omega_{\rm pe}$) in the present setup.
Electrostatic unstable modes propagating along the flow are rigorously
insensitive to the flow-aligned magnetic field. As long as $m_e/m_i\ll 1$,
the two-dimensional linear spectra computed with or without finite-mass protons are
indistinguishable. The hot spectra, accounting for the $v_{{\rm th},e}=0.01c$
thermal spread for the electrons, have then been calculated considering
infinitely heavy protons and using the fluid approximation described in 
\citet{2006PhPl...13d2106B}.

The predicted growth rates for the two-dimensional 
$k_\parallel$,$k_\perp$ wave-vector space are plotted in Figure \ref{fig:p3bret},
panels (b) and (g) for
the dilute and symmetric streams, respectively, when the stream velocity is $0.417c$ 
(corresponding to a bulk Lorentz factor $\Gamma_{\rm rel}=1.1$) {\it relative to}
the counterstream. To better accommodate the parameters available for the calculations,
it was necessary to make slight adjustments to the simulations we performed for
comparison, leading to an
increase in both the ion mass and, in the dilute case, the relative drift
speed between the plasmas. The calculations also make predictions for modes at very large
wavelengths in both spatial dimensions. We therefore repeated the early stage of our
simulation with the comparison parameters $m_i/m_e=100$ and $\Gamma_{\rm rel}=1.1$, on
an enlarged grid of $3840\Delta\times3840\Delta$, and $\lambda_{\rm se}=30\Delta$. 

To make a comparison with the analytical predictions, we extract the 
two-dimensional ${\bf k}$ spectrum at times separated by only a short interval
intended to capture the earliest linear growth of the instabilities, and compute
the average growth rate, which is plotted in Figure \ref{fig:p3bret} panels (c)--(e)
and (h)--(j) for the dilute 
and symmetric cases, respectively. The agreement is satisfactory, both qualitatively
and quantitatively, and the dominant modes are correctly rendered. As expected,
the cold theoretical spectrum saturates at high $k_\perp$ while the hot
version displays a local extremum for an oblique wave vector, as the kinetic
pressure prevents the pinching of high-$k_\perp$ small filaments 
\citep{2002PhPl....9.2458S}.

Moreover, for high $k_\perp$, the PIC spectrum describes wavelengths only
a few cells long, small enough to be affected by the smoothing algorithm. This
explains why these modes' growth is slower than expected. An electron skin depth of
several hundred cells would almost certainly provide sufficient separation from the 
filtering length for the plots to agree better with the theoretical ones in this
region, but such a simulation could not be large enough, or run long enough,
to observe the later evolution without prohibitive computational expense.

Modes with $k_\perp=0$ are purely electrostatic, and produce no magnetic
field. This is why their growth rate is much better rendered when measuring the $E$
spectrum rather than the $B$ spectrum. Conversely, modes with $k_\parallel=0$ are
mostly electromagnetic with a very small phase velocity, 
which explains why their growth is only evident in the
$B$ spectrum.

\section{Results}\label{sec:p3res}
The qualitative behavior of the counterstreaming-plasma system 
in the absence of cosmic rays exhibits a dependence on the density ratio 
$w$ particularly in the earliest stages of evolution, 
and at later times the impact of the magnetic field $b$ becomes prominent.
Although the details differ from one simulation to the next, the general behavior is
similar to that seen in the three-dimensional simulations of 
\citet{2004ApJ...608L..13F}, in which the collision of electron--ion plasmas is 
characterized by the formation of current channels first in the electrons,
and later in the ions, which proceed to merge and grow. In our experiments, 
this growth of the channels and associated structure in the magnetic field eventually reaches 
a size comparable to the simulation domain, at which point the imposed periodic boundary
conditions prevent further enlargement. However, the time necessary for this to 
occur---hundreds or thousands of $\omega_{\rm pe}^{-1}$---may exceed the 
residence time of a given particle in the subshock.

The benchmark behavior for our simulations of negligible-cosmic-ray energy density
is plotted in Figure \ref{f-na} (magnetic field absent), Figure \ref{f-nw} (weak
magnetic field), and Figure \ref{f-ns} (strong magnetic field); the three columns
of each plot correspond to the symmetric, intermediate, and dilute density ratios $w$.
The role of the density ratio is prominent in the action of the two-stream instability
on the drift velocity of the counterstreaming electron populations
\citep{1999ApJ...526..697M}. For all considered values of magnetic field $b$, 
there is an abrupt deceleration of the electrons, both the stream and counterstream, 
when $20 < \omega_{\rm pe}t < 30$.
In the $w=1$ symmetric case, this initial deceleration strips the electrons of nearly
90\% of their relative drift, but in the dilute case, less than half the drift is removed.
The electric and magnetic fields are amplified substantially
at this point, but the reduced electron drift suppresses further immediate amplification.
The ions are slower to respond, and their evolution differs qualitatively from 
that of the electrons: they form spatially alternating long-lived channels 
of current: as in lanes of vehicular traffic, 
ions moving one direction become spatially segregated from those moving the opposite direction,
and this greatly lengthens the time for the ion drift velocities to converge, though 
considerable heating of the ions takes place prior to any significant systematic deceleration.

The convergence of ion drift velocities reveals the primary effect of the magnetic field:
the ion speeds remain well separated throughout the simulation lifetime in the absent or weak 
magnetic-field cases, but the strong magnetic field brings them together within roughly 
$10^4 \omega_{\rm pe}^{-1}$ for all density ratios. It may be that at least in this
two-dimensional simulation, the transverse motion necessary for separating the ions
into current channels is slightly inhibited by the guiding magnetic field, increasing
the extent to which the counterstreaming ion populations are forced to interact. However,
it is worth noting once more that the late-term behavior of the simulations is suspect on
account of the structure size becoming comparable to the domain boundaries, an artificial
upper limit.

\subsection{Behavior including cosmic rays}\label{ssec:p3cr}
For the configurations we considered, 
the presence of cosmic rays does not appear to result in any significant deviations from
the behavior observed in their absence. When their energy density is given a 
value intended to represent conditions typical of the Galactic disk, no differences
from the negligible-cosmic-ray configuration are observed. When cosmic rays are given
an exaggerated abundance,
subtle effects do appear in the simulation, but upon inspection they are 
disregarded for one of two reasons:
they can be dismissed as numerical effects arising from the finite number of cosmic-ray 
particles employed
in our simulation, or else they result in minor quantitative changes in the evolution that
are of greatest prominence only when the effect of the periodic boundaries is already
non-negligible. When the cosmic-ray weight is comparable to that of the plasma particles,
their speed (approximately $c$) maximizes the amplitude of current-density fluctuations resulting
from the statistically expected local departures from homogeneity.

In order to verify that the observed differences resulted from our choice of representation
and not from some underlying physics, we repeated an ``abundant'' simulation with a tenfold
increase in computational particles representing the same physical cosmic-ray density.
Figure \ref{fig:p3crcp} illustrates via the electromagnetic field amplitudes 
that the statistical noise levels arising at the earliest
times saturate at a level $\sqrt{10}$ lower when cosmic rays are represented by 
50 particles per cell instead of 5, bringing both the noise level and the detailed time evolution
into better agreement with the ``negligible'' case. A further significant 
increase in the computational
particle count is too expensive for direct comparison with the simulations in Figure 
\ref{fig:p3crcp}. Using a smaller computational grid, a simulation with 500 cosmic-ray particles
per cell (leaving the other plasmas at their original 20 per cell)
illustrates the continuation of the trend observed with 50 per cell.
Nevertheless, the remaining difference in non-noise behavior is already nearly
imperceptible at just 50 computational particles per cell. This effect is paralleled in the
other aspects of the system's evolution in which abundant cosmic rays appeared to result in minor 
differences, such as drift velocities and wave spectra.

\section{Discussion and conclusions}\label{sec:p3conc}
Motivated by an interest in possible effects of cosmic rays on the physics governing
the development of collisionless astrophysical shocks,
we have performed several multidimensional
PIC simulations of
counterstreaming plasmas with various density ratios and magnetic-field strengths,
both with and without a background population of energetic cosmic rays. This
initially homogeneous environment is intended to represent the interior of the subshock,
or shock transition layer. Before cosmic
rays are added to the picture, the system resembles the subject of numerous
beam--plasma or interpenetrating-plasma studies, where the 
initial behavior of the system
can be understood in terms of known instabilities. Most prominently, the
counterstreaming electron populations are the first to interact, via
symmetric or asymmetric two-stream instabilities, as seen in, e.g.,
\citet{1999ApJ...526..697M} and \citet{2004ApJ...608L..13F}. In particular,
the three-dimensional simulations of unmagnetized electron--ion plasma collisions by
\citet{2004ApJ...608L..13F} demonstrate the formation of merging and growing 
current filaments in first the electrons and subsequently the ions, and
the nuanced relationships among the various populations.

One limitation of our approach is the reduced electron--ion mass ratio,
$m_e/m_i=1/50$. \citet{2010PhPl...17c2109B} have explored the effect of the
mass ratio on the hierarchy of unstable modes in beam--plasma systems. By
stating that a mass ratio different from 1/1836 should not change the nature
of the most unstable mode, they were able to articulate a criterion defining the
largest acceptable mass ratio. In the present case, the unstable linear spectrum
depends very weakly on the mass ratio, when no cosmic rays are introduced.
As stated in Section \ref{ssec:p3comp}, finite-mass ions do not add any extra unstable
branches to the dispersion equation. As a result, the Bret--Dieckmann
criterion is necessarily fulfilled in our configuration without cosmic rays.
Turning now to the simulations including cosmic rays, we found no significant
cosmic-ray effects with our current mass ratio within the cosmic-ray population.
It is thus likely that a real mass ratio would result in an even weaker effect,
leaving our conclusions unchanged.

Even at just 50 times the electron mass, however, the behavior of the ions
is markedly different. For the most part, the electrons produce and respond
to turbulent small-scale electromagnetic fields that serve to mix their
distribution functions; the electrons act in concert as a single population
while the ions of the stream are still easily distinguished from the ions
of the counterstream on account of the filaments. Only when the filamentary
structures have enlarged until relatively few repetitions are contained
within the simulation plane do the ions begin to converge, and in some of our
simulations that remains speculative on account of their finite duration.
While it would be possible to extend the simulations
ostensibly to observe the eventual convergence, this would be of limited value without
simultaneously increasing the dimensions of the simulation domain to mitigate the
effect of the periodic boundary conditions. An appreciable increase in size,
however, may require us to abandon the simplicity of our present configuration
by taking into account large-scale spatial variations, perhaps involving
a clear distinction between upstream and downstream regions and an unstable
charge-separation layer resulting from differences in the electrons' and ions' 
evolution.

When we compare the system evolution in the presence of cosmic rays with that
in their absence, we find that for the physical configurations studied, 
cosmic rays do not introduce a statistically significant departure from
the unperturbed results described in Section \ref{sec:p3res}. This may be a
consequence of the comparatively large mean free path and
characteristic timescale for evolution of the cosmic-ray distribution:
even with the amplification of electric and magnetic fields within the transition
layer, cosmic rays of modest energy apparently do not couple to the dynamics of thermal
electrons and ions in any appreciable way. We surmise that at least for
unmagnetized or parallel subshocks, 
the impact of cosmic rays---even when
their energy density is unusually large---on the instabilities
mediating the subshock transition is negligible.

Cosmic rays may still indirectly affect various properties of the shock, 
by modifying the upstream environment from its 
quiescent characteristics \citep{2009ApJ...706...38S}:
a shock will of necessity propagate differently through a turbulent, heated upstream
medium---perhaps with a greatly amplified magnetic field---from the comparatively clean case of a uniform, cold interstellar medium in a gently
fluctuating Galactic magnetic field.
Irrespective of the cosmic-ray abundance, 
the rapid development of turbulence in the shock transition layer and the 
associated heating of the
electrons in particular may provide an enlarged pool of candidates for injection into
the standard diffusive shock acceleration mechanism. 

The combinations of parameters we explored
did not yield any appreciable effects that could be attributed to the presence of
cosmic rays. While we chose parameters intended to be relevant to nonrelativistic
astrophysical shocks in environments where the presence of cosmic rays is suspected,
it is nevertheless possible that in other, more exotic environments, cosmic rays
may yet play some unforeseen role in the subshock microphysics. Future simulations in three
dimensions and assigning additional degrees of freedom to the magnetic field and the
cosmic-ray population may uncover effects that eluded the present analysis. However,
a departure from the quasiparallel-shock configuration will also introduce effects not
observed here, such as the development of magnetosonic waves, 
which can be more sensitive to the values
of the mass ratio or other simulation parameters; this in turn may affect the dissipation
of excited unstable modes \citep[see, e.g.,][]{2007GeoRL..3414109H}. In any case,
the dearth of differences between even the grossly exaggerated cosmic-ray energy density
and the system in which it was negligible provides a sense of reassurance that the physics
of perhaps a majority of astrophysical shock-forming instabilities can in principle be understood without
invoking some direct microscopic interference by a spectator population of cosmic rays.

This research was supported in part by the National Science Foundation
both through TeraGrid resources provided by NCSA \citep{Teragrid}  and
under Grant No. PHY05-51164. The work of A.B. is supported by projects
ENE2009-09276 of the Spanish Ministerio de Educaci\'on y Ciencia and 
PEII11-0056-1890 of the Consejer\'ia de Educaci\'on y Ciencia de la Junta
de Comunidades de Castilla-La Mancha.
The work of J.N. is supported
by MNiSW research project N N203 393034, and The Foundation for Polish
Science through the HOMING program, which is supported by a grant from
Iceland, Liechtenstein, and Norway through the EEA Financial
Mechanism. M.P. acknowledges support through grant PO 1508/1-1 of the Deutsche
Forschungsgemeinschaft.
\bibliographystyle{astron}
\bibliography{refs}

\begin{thebibliography}{}

\bibitem[\protect\astroncite{{Acciari} et~al.}{2009}]{2009Natur.462..770V}
{Acciari}, V.~A. et~al.: 2009,
\newblock {\em \nat} {\bf 462}, 770

\bibitem[\protect\astroncite{{Achterberg}}{1983}]{1983A&A...119..274A}
{Achterberg}, A.: 1983,
\newblock {\em \aap} {\bf 119}, 274

\bibitem[\protect\astroncite{{Bell}}{1978a}]{1978MNRAS.182..147B}
{Bell}, A.~R.: 1978a,
\newblock {\em \mnras} {\bf 182}, 147

\bibitem[\protect\astroncite{{Bell}}{1978b}]{1978MNRAS.182..443B}
{Bell}, A.~R.: 1978b,
\newblock {\em \mnras} {\bf 182}, 443

\bibitem[\protect\astroncite{{Bell}}{2004}]{2004MNRAS.353..550B}
{Bell}, A.~R.: 2004,
\newblock {\em \mnras} {\bf 353}, 550

\bibitem[\protect\astroncite{{Bret}}{2009}]{2009ApJ...699..990B}
{Bret}, A.: 2009,
\newblock {\em \apj} {\bf 699}, 990

\bibitem[\protect\astroncite{{Bret} and {Deutsch}}{2006}]{2006PhPl...13d2106B}
{Bret}, A. and {Deutsch}, C.: 2006,
\newblock {\em Physics of Plasmas} {\bf 13(4)}, 042106

\bibitem[\protect\astroncite{{Bret} and
  {Dieckmann}}{2010}]{2010PhPl...17c2109B}
{Bret}, A. and {Dieckmann}, M.~E.: 2010,
\newblock {\em Physics of Plasmas} {\bf 17(3)}, 032109

\bibitem[\protect\astroncite{Bret et~al.}{2010}]{bret:120501}
Bret, A., Gremillet, L., and Dieckmann, M.~E.: 2010,
\newblock {\em Physics of Plasmas} {\bf 17(12)}, 120501

\bibitem[\protect\astroncite{{Buneman}}{1993}]{1993cspp.book......M}
{Buneman}, O.: 1993,
\newblock {\em {Computer Space Plasma Physics: Simulation Techniques and
  Software, ed. H. Matsumoto \& Y. Omura}}, pp 67--84,
\newblock ~Tokyo: Terra

\bibitem[\protect\astroncite{{Catlett} et~al.}{2007}]{Teragrid}
{Catlett}, C. et~al.: 2007,
\newblock {\em {TeraGrid: Analysis of Organization, System Architecture, and
  Middleware Enabling New Types of Applications}},
\newblock IOS Press

\bibitem[\protect\astroncite{{Drury}}{1983}]{1983RPPh...46..973D}
{Drury}, L.~O.: 1983,
\newblock {\em Reports on Progress in Physics} {\bf 46}, 973

\bibitem[\protect\astroncite{{Eichler}}{1984}]{1984ApJ...277..429E}
{Eichler}, D.: 1984,
\newblock {\em \apj} {\bf 277}, 429

\bibitem[\protect\astroncite{{Frederiksen} et~al.}{2004}]{2004ApJ...608L..13F}
{Frederiksen}, J.~T., {Hededal}, C.~B., {Haugb{\o}lle}, T., and {Nordlund},
  {\AA}.: 2004,
\newblock {\em \apjl} {\bf 608}, L13

\bibitem[\protect\astroncite{{Gargat{\'e}} et~al.}{2010}]{2010ApJ...711L.127G}
{Gargat{\'e}}, L., {Fonseca}, R.~A., {Niemiec}, J., {Pohl}, M., {Bingham}, R.,
  and {Silva}, L.~O.: 2010,
\newblock {\em \apjl} {\bf 711}, L127

\bibitem[\protect\astroncite{{Greenwood} et~al.}{2004}]{2004JCoPh.201..665G}
{Greenwood}, A.~D., {Cartwright}, K.~L., {Luginsland}, J.~W., and {Baca},
  E.~A.: 2004,
\newblock {\em Journal of Computational Physics} {\bf 201}, 665

\bibitem[\protect\astroncite{{Hellinger} et~al.}{2007}]{2007GeoRL..3414109H}
{Hellinger}, P., {Tr{\'a}vn{\'{\i}}{\v c}ek}, P., {Lemb{\`e}ge}, B., and
  {Savoini}, P.: 2007,
\newblock {\em \grl} {\bf 34}, L14109

\bibitem[\protect\astroncite{{Lemoine} and
  {Pelletier}}{2011}]{2011MNRAS.417.1148L}
{Lemoine}, M. and {Pelletier}, G.: 2011,
\newblock {\em \mnras} {\bf 417}, 1148

\bibitem[\protect\astroncite{{Luo} and {Melrose}}{2009}]{2009MNRAS.397.1402L}
{Luo}, Q. and {Melrose}, D.: 2009,
\newblock {\em \mnras} {\bf 397}, 1402

\bibitem[\protect\astroncite{{Malkov} and {Drury}}{2001}]{2001RPPh...64..429M}
{Malkov}, M.~A. and {Drury}, L.~O.: 2001,
\newblock {\em Reports on Progress in Physics} {\bf 64}, 429

\bibitem[\protect\astroncite{{Medvedev} and {Loeb}}{1999}]{1999ApJ...526..697M}
{Medvedev}, M.~V. and {Loeb}, A.: 1999,
\newblock {\em \apj} {\bf 526}, 697

\bibitem[\protect\astroncite{{Nakar} et~al.}{2011}]{2011ApJ...738...93N}
{Nakar}, E., {Bret}, A., and {Milosavljevi{\'c}}, M.: 2011,
\newblock {\em \apj} {\bf 738}, 93

\bibitem[\protect\astroncite{{Niemiec} et~al.}{2008}]{2008ApJ...684.1174N}
{Niemiec}, J., {Pohl}, M., {Stroman}, T., and {Nishikawa}, K.: 2008,
\newblock {\em \apj} {\bf 684}, 1174

\bibitem[\protect\astroncite{{Ohira} et~al.}{2009}]{2009ApJ...698..445O}
{Ohira}, Y., {Reville}, B., {Kirk}, J.~G., and {Takahara}, F.: 2009,
\newblock {\em \apj} {\bf 698}, 445

\bibitem[\protect\astroncite{{Rabinak} et~al.}{2011}]{2011ApJ...736..157R}
{Rabinak}, I., {Katz}, B., and {Waxman}, E.: 2011,
\newblock {\em \apj} {\bf 736}, 157

\bibitem[\protect\astroncite{{Rakowski} et~al.}{2008}]{2008ApJ...684..348R}
{Rakowski}, C.~E., {Laming}, J.~M., and {Ghavamian}, P.: 2008,
\newblock {\em \apj} {\bf 684}, 348

\bibitem[\protect\astroncite{{Reville} et~al.}{2006}]{2006PPCF...48.1741R}
{Reville}, B., {Kirk}, J.~G., and {Duffy}, P.: 2006,
\newblock {\em Plasma Physics and Controlled Fusion} {\bf 48}, 1741

\bibitem[\protect\astroncite{{Reynolds}}{2008}]{2008ARA&A..46...89R}
{Reynolds}, S.~P.: 2008,
\newblock {\em \araa} {\bf 46}, 89

\bibitem[\protect\astroncite{{Riquelme} and
  {Spitkovsky}}{2009}]{2009ApJ...694..626R}
{Riquelme}, M.~A. and {Spitkovsky}, A.: 2009,
\newblock {\em \apj} {\bf 694}, 626

\bibitem[\protect\astroncite{{Scholer} et~al.}{2002}]{2002PhPl....9.4293S}
{Scholer}, M., {Kucharek}, H., and {Kato}, C.: 2002,
\newblock {\em Physics of Plasmas} {\bf 9}, 4293

\bibitem[\protect\astroncite{{Silva} et~al.}{2003}]{2003ApJ...596L.121S}
{Silva}, L.~O., {Fonseca}, R.~A., {Tonge}, J.~W., {Dawson}, J.~M., {Mori},
  W.~B., and {Medvedev}, M.~V.: 2003,
\newblock {\em \apjl} {\bf 596}, L121

\bibitem[\protect\astroncite{{Silva} et~al.}{2002}]{2002PhPl....9.2458S}
{Silva}, L.~O., {Fonseca}, R.~A., {Tonge}, J.~W., {Mori}, W.~B., and {Dawson},
  J.~M.: 2002,
\newblock {\em Physics of Plasmas} {\bf 9}, 2458

\bibitem[\protect\astroncite{{Stroman} et~al.}{2009}]{2009ApJ...706...38S}
{Stroman}, T., {Pohl}, M., and {Niemiec}, J.: 2009,
\newblock {\em \apj} {\bf 706}, 38

\bibitem[\protect\astroncite{{Umeda} et~al.}{2003}]{2003CoPhC.156...73U}
{Umeda}, T., {Omura}, Y., {Tominaga}, T., and {Matsumoto}, H.: 2003,
\newblock {\em Computer Physics Communications} {\bf 156}, 73

\bibitem[\protect\astroncite{{Webb} et~al.}{1983}]{1983ApJ...270..537W}
{Webb}, G.~M., {Axford}, W.~I., and {Terasawa}, T.: 1983,
\newblock {\em \apj} {\bf 270}, 537

\end{thebibliography}
\begin{figure}[p]

\includegraphics[width=6.2in]{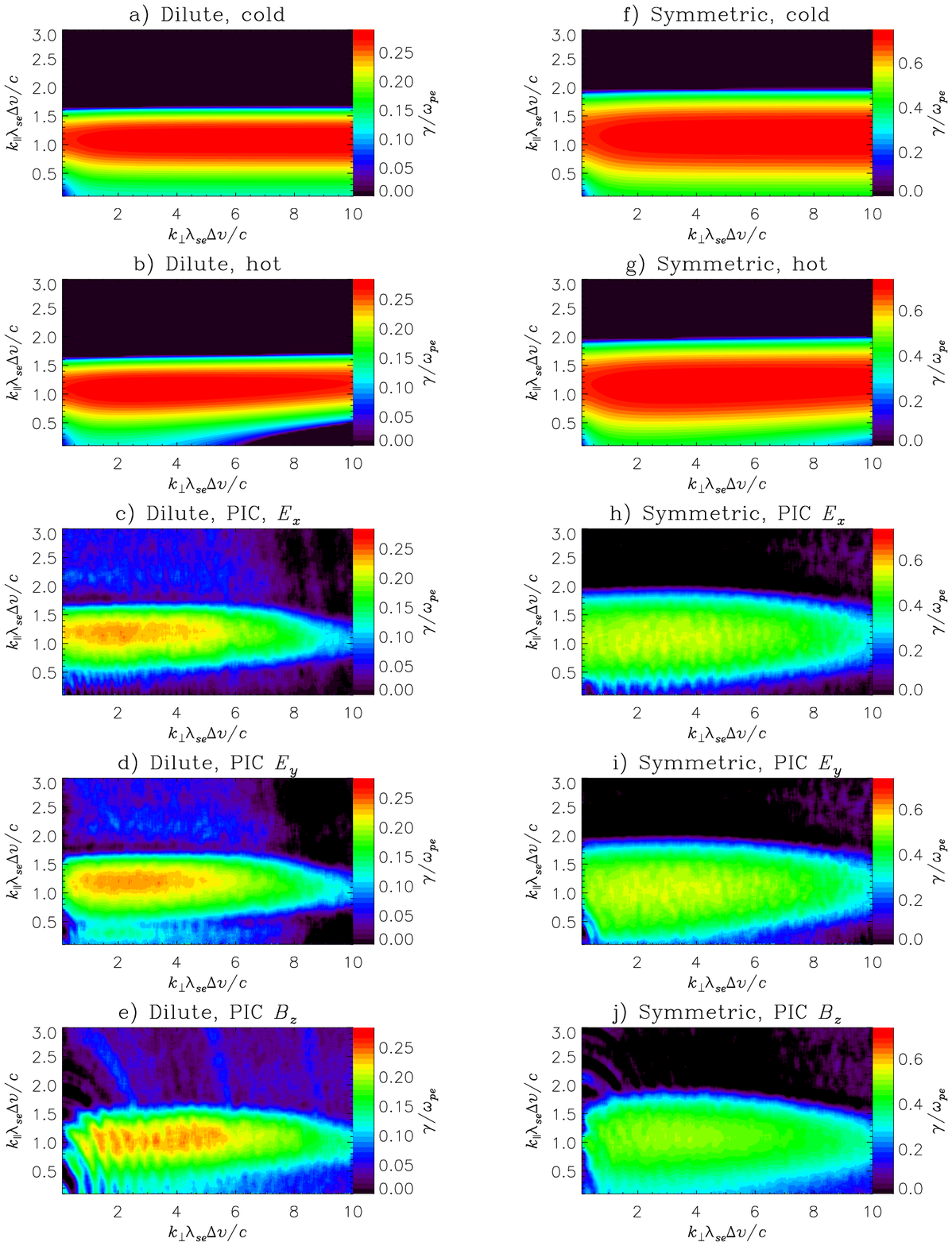}

\end{figure}
\begin{figure}[p]
\caption[Growth rates of early unstable modes]{Measured or calculated
growth rates (in units of electron plasma frequency) 
of unstable modes with wave vector $k_\parallel,k_\perp$ 
(scaled by 
$\lambda_{\rm se}\Delta v/c$) for
the dilute (left column) and symmetric (right column) stream configurations, 
with negligible-cosmic-ray density and the weak ($b=0.01$) magnetic field.
Within each column, the uppermost plot (a, f) is the instantaneous growth rate
calculated in the zero-temperature limit. The next plot (b, g) includes a finite thermal
spread of $v_{{\rm th},e}=0.01c$ in the calculation. The remaining
three rows are the average growth rate measured in the early stages of 
high-resolution PIC simulations, for the growth of perturbations in $E_x$ (c, h),
$E_y$ (d, i), and $B_z$ (e, j). The growth-rate measurement is
performed by calculating the fractional increase in the
instantaneous two-dimensional Fourier power spectrum
for the selected field component between times $t_0$ and $t_1$.
For the dilute configuration, $\omega_{\rm pe}t_0\approx 6$ and $\omega_{\rm pe}t_1\approx 18$;
for the symmetric configuration, $\omega_{\rm pe}t_0\approx 6$ and $\omega_{\rm pe}t_1\approx 12$.
By virtue of being an average over a finite interval, 
the growth rate obtained in this way is not a pure,
instantaneous quantity like the top two rows of plots. The changing conditions
in the stream and counterstream alter the instantaneous growth rate during the
measurement interval, resulting in the minor differences that appear between
the upper and lower parts of each column.
}
\label{fig:p3bret}
\end{figure}

\begin{figure}[p]

\includegraphics[width=6in]{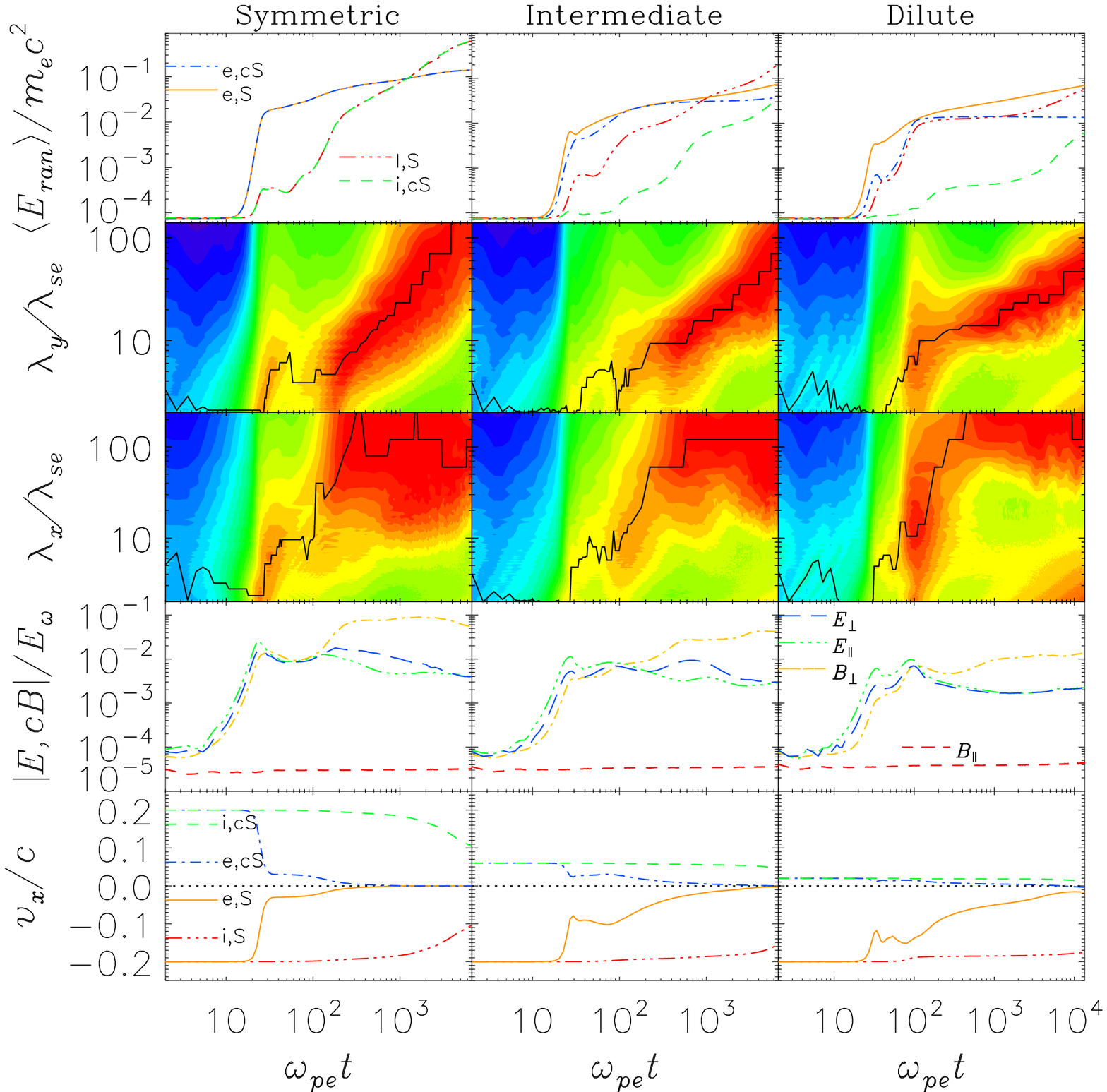}
\end{figure}
\begin{figure}[p]
\caption[Selected results from the ``negligible-absent'' family of 
simulations]{Selected results from the ``negligible-absent'' family of simulations,
in which cosmic rays are negligible and an initial homogeneous magnetic field is absent. From
left to right, the density ratio of the stream to the counterstream is $w=1$, 0.3, and 0.1, 
corresponding to the ``symmetric,'' ``intermediate,'' and ``dilute'' cases, respectively.
The bottom row shows the drift velocity of the stream electrons (solid line), stream ions 
(dot-dot-dot-dashed line), counterstream electrons (dot-dashed line), and counterstream ions
(dashed line). A dotted line is included at zero velocity, representing the cosmic-ray ions
and electrons. The second-to-bottom row displays the root-mean-square amplitudes of the electric
and magnetic field parallel and perpendicular to the drift axis, where 
the magnetic field has been multiplied by $c$ and all components are expressed in units of 
$E_\omega\equiv\omega_{\rm pe}\,c\,m_e/e$. 
The second and third rows from the top illustrate the time
evolution of the Fourier spectra of $B_z$ both parallel and perpendicular to the drift, with  
wavelengths measured in units of the electron skin depth $\lambda_{\rm se}$. A solid black line traces
the evolution of the dominant values of $\lambda_x$ and $\lambda_y$. The top row plots the mean
random kinetic energy (i.e., thermal energy) per particle, $E_{\rm ran}=\left(\gamma'-1\right)\,m\,c^2$, 
where $\gamma'$ is the particle Lorentz factor in the local plasma rest frame (measured over a
square-shaped region of area $100 \Delta^2$). The lines in the top row represent the same 
populations as in the bottom row.}
\label{f-na}
\end{figure}

\begin{figure}[p]

\includegraphics[width=6in]{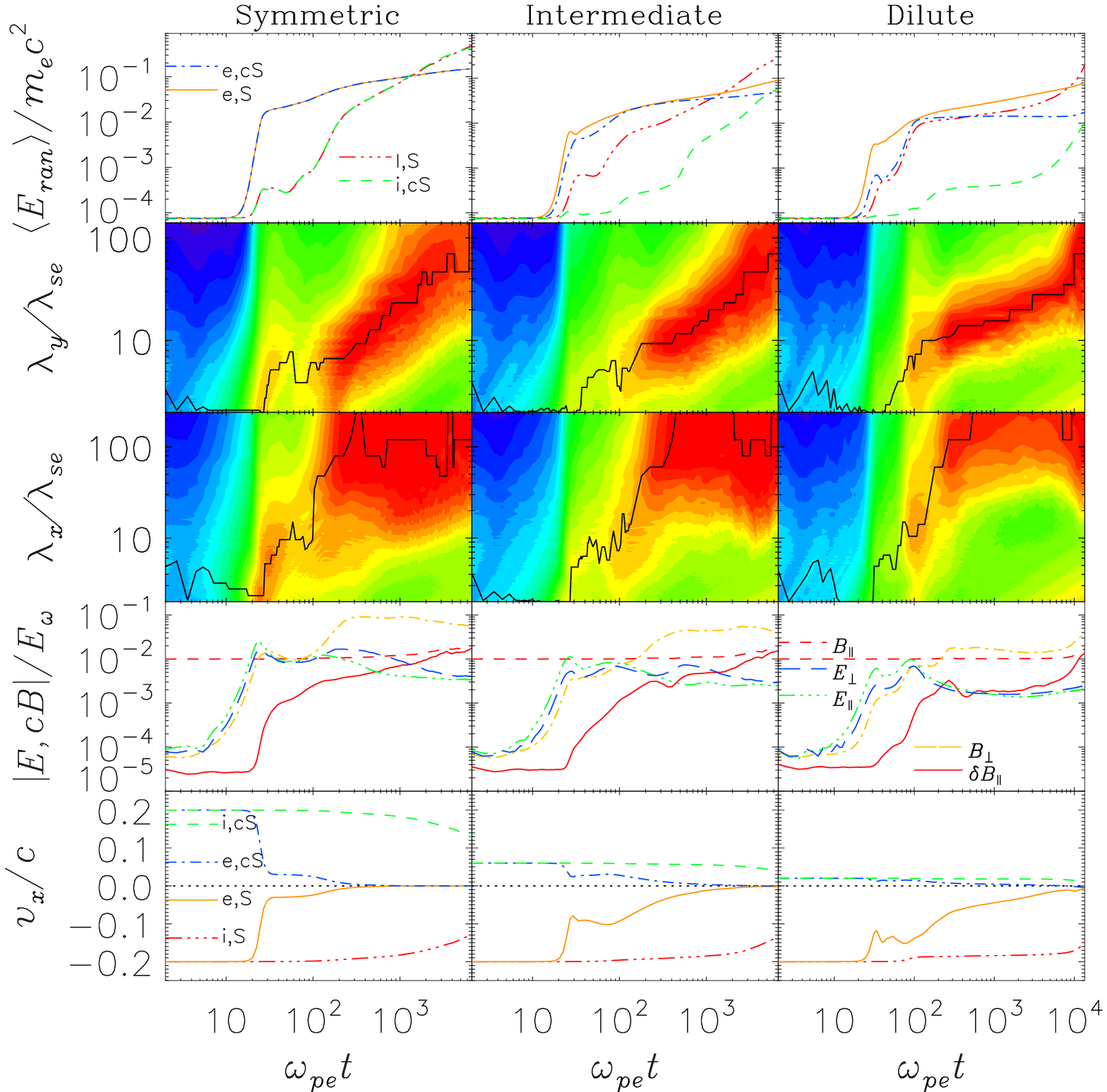}
\caption[Selected results from the ``negligible-weak'' family of 
simulations]{Selected 
results from the ``negligible-weak'' family of simulations. The initial magnetic field is set such that
the electron cyclotron frequency is $\Omega_e\equiv e\,B/m_e=0.01\omega_{\rm pe}$. See the caption
to Figure \ref{f-na} for a detailed description of each plot.}
\label{f-nw}
\end{figure}

\begin{figure}[p]

\includegraphics[width=6in]{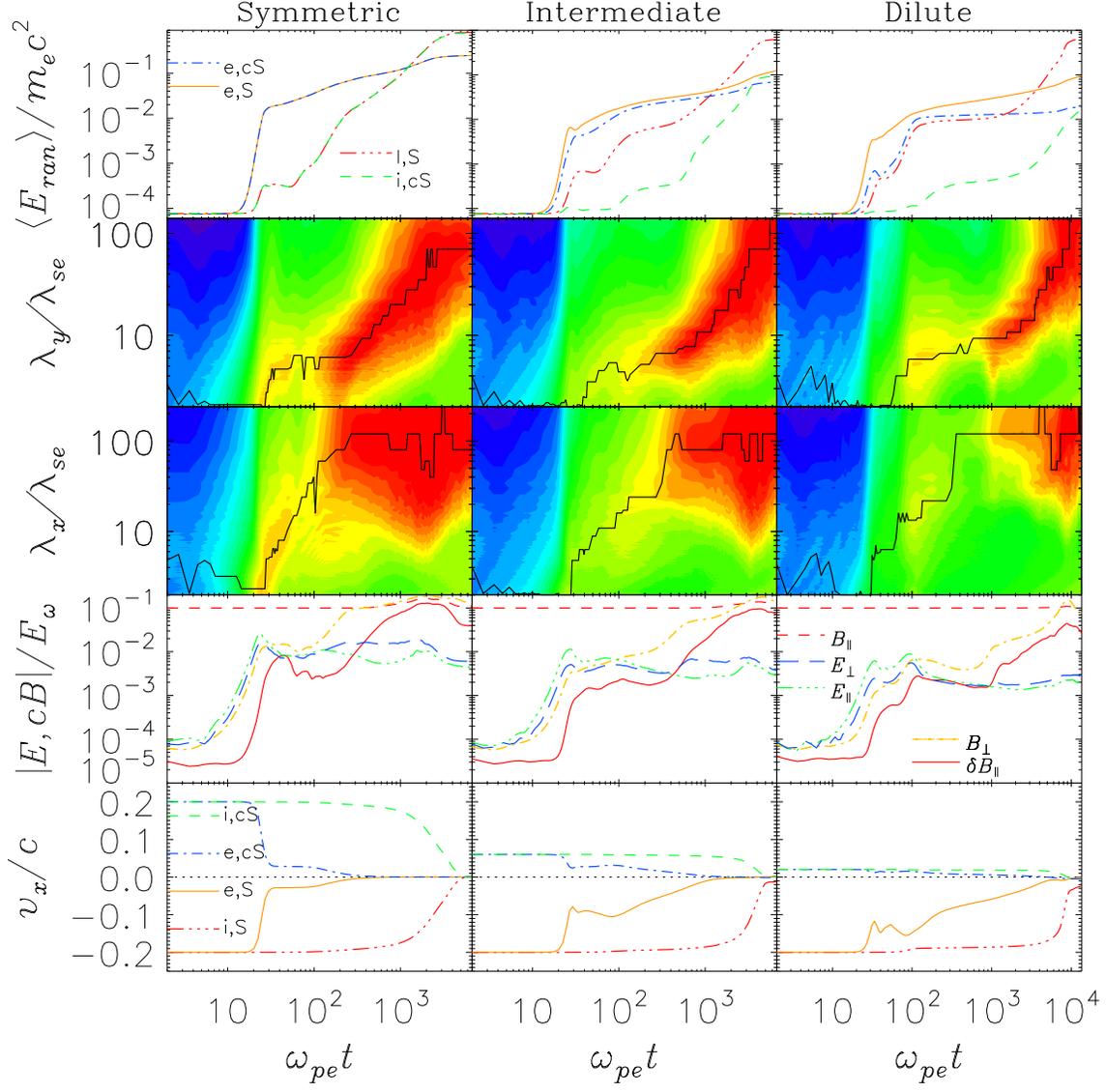}
\caption[Selected results from the ``negligible-strong'' family of 
simulations]{Selected results from the ``negligible-strong'' family of simulations. The 
initial magnetic field is set such that
the electron cyclotron frequency is $\Omega_e=0.1\omega_{\rm pe}$. See the caption
to Figure \ref{f-na} for a detailed description of each plot.}
\label{f-ns}
\end{figure}

\begin{figure}[p]
\includegraphics[width=6.2in]{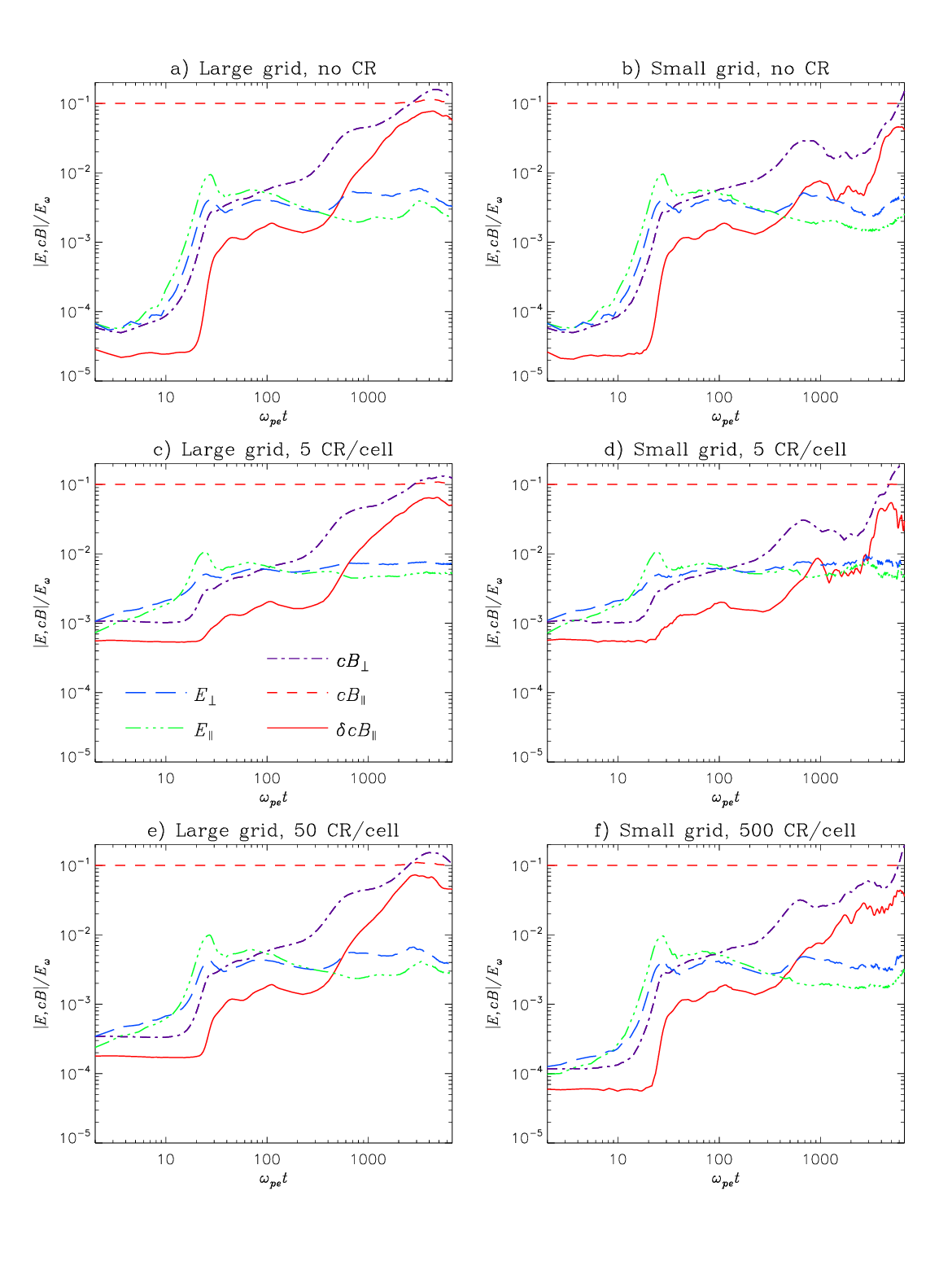}
\end{figure}
\begin{figure}[h]
\caption[Cosmic-ray representation effects]
{Evolution of the electric and magnetic field components in
the configuration $b=0.1$, $w=0.3$, as a function of the cosmic rays' computational
representation. Panel (a) shows the case when cosmic rays are negligible.
The ``abundant'' ($w_{\rm CR}\gamma_{\rm CR}=10$) cosmic-ray distribution is represented by only five 
computational particles per cell in panel (c). In panel (e), the number of particles has been
increased tenfold, and the initial field amplitudes decrease by approximately $\sqrt{10}$,
suggesting that statistical fluctuations in the cosmic-ray contribution to the local
current density result in a high noise level in the field components. The column on the
right reproduces the conditions of the left column on a grid whose dimensions have been
reduced by 90\% in each direction, allowing a full-length simulation with 500 cosmic-ray
particles per cell. Panels (b) and (d) are comparable to panels 
(a) and (c), respectively, until $\omega_{\rm pe}t\sim 1000$; the difference
beyond this is a result of the smaller
grid's periodic boundaries inhibiting the continued spatial growth of structures earlier.
In panel (f), the initial noise level is reduced by a factor of $\sqrt{100}$ in
all field components, corroborating our earlier supposition that the higher amplitudes
are the effect of statistical noise.
}
\label{fig:p3crcp}
\end{figure}
\end{document}